# Terahertz metasurface sensor with graphene microstrips for biosensing: modeling and application

K. S. Kuznetsova[1], V. A. Pashynska[1,2], Z. E. Eremenko[1,3]

[1] O.Ya. Usikov Institute for Radiophysics and Electronics, NAS of Ukraine, 12 Academ. Proskury Str., Kharkiv 61085, Ukraine, kuznetsova.ire@gmail.com

[2] B.Verkin Institute for Low Temperature Physics and Engineering of NAS of Ukraine, 47 Nauky Ave., Kharkiv, 61103, Ukraine, vlada.pashynska@gmail.com

[3] Leibniz Institute for Solid State and Materials Research, 01069 Dresden, Germany, zoya.eremenko@gmail.com

**Abstract.** The present numerical study is devoted to development and optimization of a metasurface-based sensor with graphene constituents for potential biosensing applications. A unit cell of the proposed metasurface consists of a thin flexible dielectric substrate layer with a centrally positioned graphene microstrip. As a result of numerical modeling of spectral properties of the metasurface by COMSOL Multiphysics software in terahertz range from 5 to 35 THz the absorption spectrum maxima (resonance modes) at $f_1$ = 8.7 THz and $f_2$ = 26.5 are revealed. Structural parameters of the developed metasurface with graphene microstrips have been tuned to achieve the optimal resonance properties. The following stages of the study demonstrate that placement of a layer of tested liquid sample (water or Bovine Serum Albumin (BSA) solution) on the metasurface causes a low frequency shift of the plasmonic resonance mode $f_1$ chosen for biosensing measurements. This frequency shift, along with the change in the amplitude of the absorption peak, are highly sensitive to the refractive index of the tested liquid sample. The resonance behavior of the developed metasurface structure is governed by the excitation of localized plasmon resonance in the graphene elements and near-field electromagnetic coupling effect between the short edges of the graphene microstrips. Evaluation of the influence of dielectric substrate's material on the sensitivity of the metasurface-based sensor to variations in BSA concentration indicates that Kapton substrate provides higher effectiveness compared with SiC substrate. The obtained results demonstrate the potential of the developed metasurface-based sensor with graphene microstrips for application as a sensing structure to determine proteins and other biomolecules in liquid samples.

# 1. INTRODUCTION

Graphene is considered as promising nanomaterial for biosensing applications due to its biocompatibility and exceptional physicochemical properties, including electromagnetic and optical characteristics that are well-suited for sensing [1–3]. Graphene micro- and nanostrips used as sensing elements in graphene based biosensors can be produced by several proven methods including sonochemical method, lithography, chemical vapour deposition, plasma etching, lithium intercalation/exfoliation of carbon nanotubes, and others [4]. Graphene sensors offer several advantages, such as rapid photoresponse, broadband photodetection (from the ultraviolet to the terahertz (THz) spectral regions), high sensitivity, and reasonable cost [3]. The modern sensing devices based on graphene demonstrate significant potential in medical diagnostics, drug discovery, food safety analysis, and biological reactions monitoring [1].

Among the various proposed graphene-based sensor architectures, metasurfaces incorporating graphene have gained considerable attention in recent years. This is due to their unique and tunable optical and electronic properties, which allow precise control over plasmonic resonances in such structures. These features create exciting opportunities for the development of sensors, absorbers, and other devices [1, 5, 6] on the basis of such metasurfaces. Precise control of surface plasmon interactions provided by structural modifications of graphene-based metasurfaces is essential for optimizing device performance. The unique plasmonic properties of graphene have established it as a cornerstone material in contemporary biological sensing research, driving innovations in ultra-sensitive detection platforms. Recently developed graphene-based metasurfaces have demonstrated their effectiveness as sensing platforms for the detection of diagnostically significant biomolecules, such as DNA and proteins (including antibodies and enzymes), in biological samples [1].

It is important to note that currently the analytical biochemical methods (ELISA tests, dyes applications etc.) prevail in biomedical practice for determination of biologically significant proteins (diagnostic markers, antibodies etc.) in biological liquid samples. The ELISA tests are effective, but rather expensive in development due to their high specificity and necessity of special biochemical laboratory equipment and chemical reagents for application. That is why the development of robust physical methods (which do not require the use of antibodies and reagents) for the detection of proteins in liquid samples remains demanded. Specifically, study of plasmonic effects in the metasurfaces with graphene sensing components for development of biosensors on their basis can be considered as an actual task of modern biomedical related nanotechnology and condensed matter science areas.

The present work aims to design and numerically model a metasurface sensing structure that incorporates graphene microstrips embedded within a dielectric layer, with a superimposed layer of tested liquid analyte. As an analyte layer placed above the dielectric substrate, we use water and aqueous solutions of Bovine Serum Albumin (BSA). It should be noted that albumin level measurements are important for medical diagnostic purposes because of serum albumin is the most abundant protein in human blood plasma and plays a number of critical regulating roles in maintaining the physiological balance [7]. Abnormal albumin levels can indicate a wide range of health disorders in humans and animals [8]. Owing to the presence of collective vibrational and torsional modes in the THz range, proteins (including albumin) can be effectively characterized at the molecular level using THz spectroscopy [9]. In particular, it has been established that an increase in BSA concentration in the solution leads to a consistent decrease of spectral absorption intensity, which is associated with changes in the dielectric properties of the liquid sample [10].

In our previous studies on the development of the sensor structures based on the dielectric and metal-dielectric metasurfaces [11, 12] we proposed and tested the approach that includes using our experimental differential dielectrometry data of complex permittivity for biologically important proteins solutions [13] in the numerical modeling with COMSOL software. This approach is employed in the present study to evaluate the performance of the developed metasurface containing graphene microstrips for BSA level determination in aqueous solutions.

The main objective of the current study is to reveal how the tested liquid's physical properties (in particular, the refractive index) affect the resonance frequency and other spectral characteristics of the graphene-based metasurface which is important for the prospect of such structures applications for biosensing.

## 2. NUMERICAL MODELING METHODOLOGY

Numerical modeling in the current study was performed using COMSOL Multiphysics software (O.Ya. Usykov Institute for Radiophysics and Electronics, NAS of Ukraine (IRE) license No. 17078683), taking into account our previous experience in the development of biosensing structures [11, 12]. Simulations were conducted in THz range by variation of the refractive index values of the tested liquid samples. Periodic Floquet boundary conditions were applied along the x- and y- axes. In the simulation, the electric field vector is oriented along the y-axis. The Kubo formula was used to define graphene's surface conductivity [14]. In this work, the Fermi level of graphene is set to $Ef$=0.5eV.

Key parameters studied within the modeling include the absorption spectrum, resonance shift, and electric field distribution near the graphene strip at resonance conditions.

## 3. RESULTS AND DISCUSSION

### 3.1. Metasurface unit cell design

In the present study we propose the design of the metasurface with graphene components, the unit cell of which is presented in Fig. 1. The unit cell of the metasurface consists of a thin dielectric substrate (thickness td = 1.25 μm, corresponding 0.06 of the wavelength in the dielectric substrate at the resonance frequency $f_1$ of the blank metasurface) with a centrally positioned graphene microstrip with the short and long edges of size W = 0.4 μm, L = 6.36 μm, respectively; thickness of the microstrip is 0.1 μm (Fig. 1). The unit cell structure exhibits two different periodicities along the x- and y-axes, denoted as $P_1$ = 2.5 μm and $P_2$ = 7 μm, respectively. In the proposed model, Kapton is used as a dielectric substrate due to its low dielectric losses ($\varepsilon'$ = 3.28, $\varepsilon''$ = 0.056) so that the overall absorption spectra characteristics of the metasurface are not influenced by the substrate material. In the research [14], it was shown that the compatibility of Kapton with graphene improves the sensor tuning and absorption characteristics in biosensing. A gold layer is placed beneath the dielectric in the proposed metasurface to serve as a reflective backing for elimination of transmission and enhancing the field confinement within the structure. Unlike conventional configurations where graphene elements are deposited on the surface of the dielectric substrate, in our model, graphene microstrips are embedded within the dielectric bulk at a depth of 0.25 μm from the surface, thus limiting its direct interaction with biological analytes.

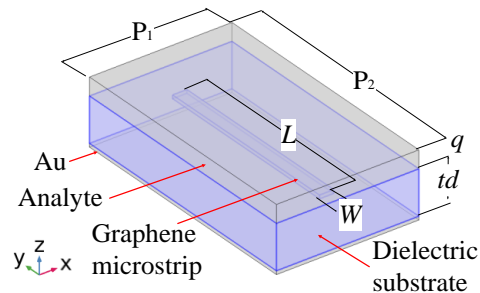

**Fig. 1.** Unit cell of the developed graphene-based metasurface consisting of dielectric layer with a centrally positioned graphene microstrip and tested liquid layer of thickness q=0.5 μm.

### 3.2. Spectral characterization and structural tuning of the metasurface

The simulation results demonstrate that the graphene-based metasurface exhibits two resonance modes in THz range at $f_1$ = 8.7 THz and $f_2$ = 26.5 THz (Fig. 2a). These modes are characterized by similar absorption peak intensities, with the absorption calculated as $A(f)=1-|S_{11}(f)|^2$. The normalized electric field distribution for the two resonant modes differs significantly, as shown in Fig. 2b. For $f_1$ mode, strong field localization is observed in the gaps between the short edges W of the adjacent graphene microstrips (fundamental dipole mode). This indicates significant near-field coupling between neighboring graphene elements of the metasurface in which the distance between graphene microstrips is significantly less than the radiation wavelength. In contrast, $f_2$ mode shows field enhancement not only at the edges, but also along the entire surface of the graphene microstrips in the form of two scattered spots, which are standing waves along the longer sides L of the microstrips [16]. Both resonant modes exhibit electric field distributions similar to plasmonic resonances in graphene ribbons, especially near the edges of the graphene microstructures [16-18]. We can classify these two resonances as a dipole mode ($f_1$) and a lattice mode ($f_2$). These modes are of the same type as those observed in metallic or all-dielectric metasurfaces [19, 20], namely dipole-like and collective lattice resonances. While supporting the same fundamental types of the resonance modes as in metallic or dielectric based metasurfaces, graphene-based structures provide an additional advantage — electrical tunability. This enables reconfigurable and application-oriented operation graphene-based sensors in the terahertz range. The field profiles of our graphene-based metasurface are presented in the top and bottom panels of Fig. 2b for the $f_1$ and $f_2$ modes, respectively.

The localized plasmon mode $f_1$ exhibits a dipole-like field distribution and shows high sensitivity to variations on the geometric parameters of the metasurface unit cell. Such sensitivity, as shown below, enables the metasurface effective application for determination of the protein concentration in liquid samples using the resonance frequency shift of this plasmon mode. Taking into account the mentioned above characteristics of the surface plasmon mode $f_1$ and based on the known correspondence of water-related vibrational modes to the lower part of the THz spectrum [21], we chosen this first resonance mode $f_1$ for examining BSA solutions of different concentration. Moreover, the spectral region of $f_1$ contributes more to the characteristic absorption spectra observed in THz range for aqueous biomolecular solutions, such as BSA and thus provides high sensitivity to the protein concentration.

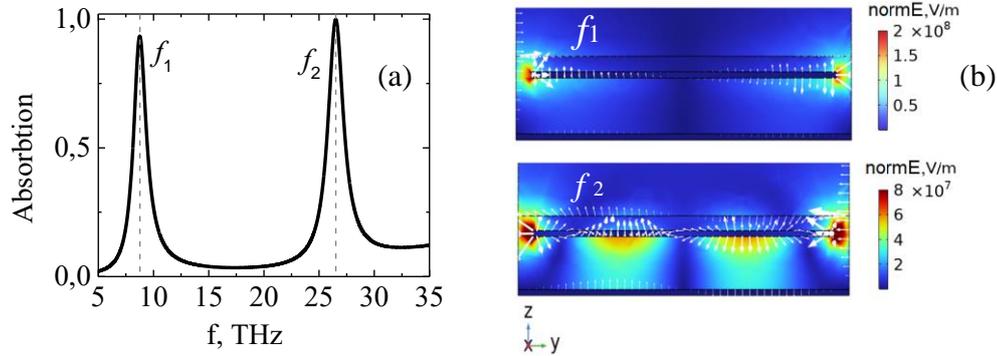

**Fig. 2.** (a) Absorption spectra of the metasurface; (b) Electric field distribution in the cross section along the side $P_2$ of the metasurface unit cell at resonance frequencies without analyte.

Figure 3 illustrates the tunability of the proposed metasurface resonance properties against variations of geometric parameters of the unit cell, including the graphene microstrip length L (a), width W (b), and thickness of the dielectric substrate td (c). The most pronounced tuning of the resonance frequency in the absorption spectrum of the metasurface can be achieved by varying the length L of the graphene strip (Fig. 2a). Increasing or decreasing the strip length changes the charge oscillation path, directly shifting the plasmonic resonance frequency. Additionally, the near-field electromagnetic coupling between the short edges W of the microstrips is highly sensitive to the distance between these edges, which is also determined by the microstrip length L. These effects lead to a more significant dependency of the resonance frequency of the metasurface from L value compared to changes of other structural parameters. The results of our modeling show that the highest absorption intensity is observed when the length and width of the graphene microstrips are L = 6.36 μm and W = 0.4 μm, respectively, with the dielectric substrate thickness td = 1.25 μm.

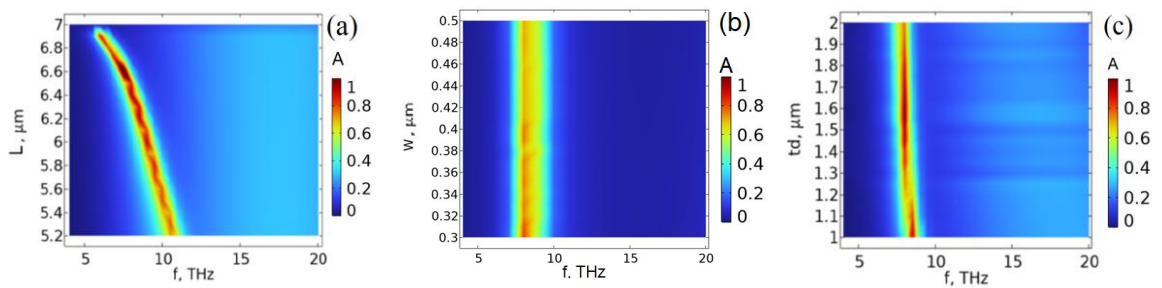

Fig. 3. Absorption spectra of the metasurface without analyte under variation of different geometric parameters of the unit cell: (a) graphene microstrip length L variation; (b) graphene microstrip width W variation at fixed L = 6.36 µm; (c) dielectric substrate thickness td variation at L = 6.36 µm.

### 3.3. Quantitative determination of BSA in the solution using developed sensor structure

To detect and evaluate the BSA content in the liquid samples, we selected the frequency range of 4-12 THz for the designed metasurface, operating on the electric dipole resonant mode $f_1$. We tuned the frequency and amplitude of the plasmonic resonance $f_1$ by varying the structural parameters of the unit cell of the metasurface with the graphene microstrip. In the simulations, BSA solutions were defined via their refractive index $n$, which was varied according to the corresponding concentrations of the tested solutions.

Figure 4 presents the absorption spectra of the metasurface with a liquid analyte layer (thickness q = 0.5 µm) deposited on the surface and with usage of two different dielectric substrates: Kapton (a) and SiC (b), for comparison. The resonances are observed at 8.7 THz and 5 THz for the bare metasurface with Kapton and SiC dielectric substrates, respectively. Upon the addition of water, the resonance shifts to 7.4 THz for the Kapton-based structure, and shifts to 4.8 THz for the SiC-based one. The modeling was performed for the BSA aqueous solutions within the concentration diapason 0-200 mg/mL (it should be noted that the physiological concentration of albumin in human serum is in the range of 35-55 mg/mL for adults, the concentration of albumin in technological solutions can be higher depending on the application purposes). For pure water and a low BSA concentration of 10 mg/mL the spectral absorption response remains almost unchanged (Fig. 4, a, b). The observed broadening of the resonance peak with liquid analyte loading suggests increased losses, caused by the absorption properties of water and aqueous solutions of the protein. With the following growing of the concentration of BSA in the aqueous solutions, a noticeable decrease in absorption peak intensity is observed, along with a shift of the resonance frequency to the low frequency region (see Fig. 4 a, b).

We revealed that the resonance frequency shift $\Delta f$ is bigger, and therefore the sensitivity to BSA concentration changes is higher for the sensor structure containing Kapton dielectric substrate (Fig. 4, c, curve 1) compared to SiC (Fig. 4, c, curve 2). To precisely describe these dependencies, the simulation data were fitted with second-order polynomial equations. The analysis demonstrates how the choice of dielectric substrate material influences the resonance behaviour and absorption spectra characteristics of the metasurface in the presence of the analyte.

A quantitative evaluation of the performance of the proposed sensor containing Kapton was carried out based on the sensitivity S = Δ$f$/Δ$n$ [THz/RIU] to BSA detection, which amounted to 1.6 THz/RIU. The sensitivity of the proposed graphene metasurface sensor for BSA determination is comparable to or higher than that of other reported THz sensors. For example, a dielectric-based metamaterial sensor demonstrated a sensitivity of 3.2 THz/RIU for virus detection [22], while another graphene-based metasurface operating in the 2–6 THz band achieved 0.85 THz/RIU when varying the refractive index of the test medium [23]. A metal-dielectric metamaterial biosensor for BSA reported a significantly lower sensitivity of 0.135 THz/RIU [24].

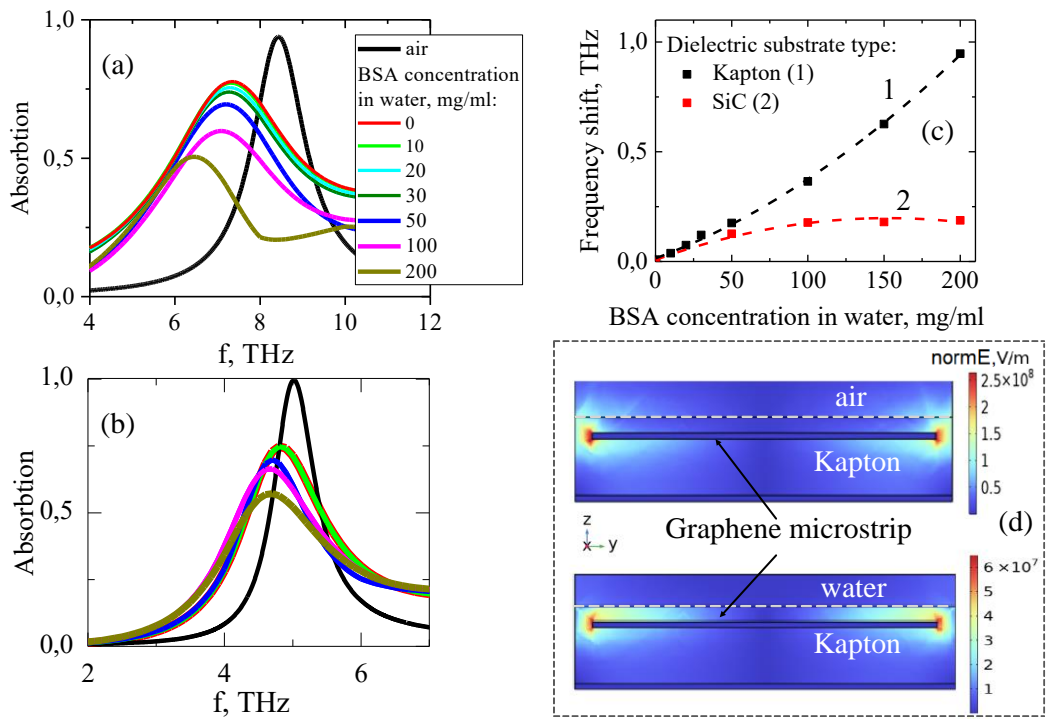

Fig. 4. Absorption spectrum of the metasurface-based sensor with graphene microstrips for two types of dielectric substrate such as Kapton (a) and SiC (b) for air and different BSA concentrations in aqueous solutions. (c) The resonance frequency shift for the metasurface depending on analyte concentration for two types of dielectric substrate. (d) Electric field distribution in the cross section of the unit cell of the metasurface without analyte (air) and with analyte (water) on the metasurface. The white dashed line marks the separation between the dielectric substrate layer with the graphene strip and location of the analyte.

The electric field distributions were analyzed to explain the absorption spectra behavior of the metasurface used for the biosensing. Fig. 4 (d) presents the normalized electric field distribution in cross section along the long side P$_2$ for the metasurface unit cell at $f_1$ resonance frequency.

Without the analyte (air above the structure), the electric field extends beyond the dielectric structure with the graphene microstrip into the surrounding environment. Localized maxima remain inside the structure near the edges of the graphene element. Under the analyte presence (the layer of water), the electric field is limited within the dielectric structure containing the graphene microstrip. When the analyte is uploaded on the surface, the field becomes largely confined within the structure (bottom panel of Fig. 4 d), which in turn influences the intensity of the absorption peak (Fig. 4 a). This enhanced field localization can be exploited for the detection of biologically active substances in liquid samples, as even slight changes in the refractive index of the analyte can noticeably affect the resonance behavior.

## 4. CONCLUSIONS

In the current modeling study the design of metasurface contained graphene microstrips as sensing elements was developed and optimized for potential biosensing applications. Numerical modeling using COMSOL Multiphysics software in terahertz range demonstrated that the developed metasurface-based sensing structure with graphene elements has two resonance modes at $f_1$ = 8.7 THz and $f_2$ = 26.5 THz. It was shown that variations of geometrical characteristics of the cell unit of the metasurface, such as the microstrip length, width, and the thickness of the dielectric substrate, significantly influence the resonant properties of the sensing structure. The following optimal unit cell parameters were identified: the length and width of the graphene microstrips were L = 6.36 µm and W = 0.4 µm, respectively, dielectric substrate thickness - td = 1.25 µm.

In the subsequent stage of our study it was revealed that placement of a layer of the tested liquid sample (water or Bovine Serum Albumin solution) to the metasurface resulted in a shift of the plasmonic resonance mode $f_1$ toward lower frequencies and in a conspicuous change in the absorption peak intensity. The resonant behavior of the sensor was examined at different concentrations of the tested BSA solutions (in the range from 0 to 200 mg/ml) imposed on the metasurface and for different dielectric substrate materials (Kapton and SiC). The developed graphene microstrips containing metasurface demonstrated reliable sensitivity to the changes of the BSA concentration in the solutions tested. Sensing metasurface with Kapton substrate shown the higher effectiveness compared to the metasurface with SiC.

The spatial distribution of the electric field at resonance within the unit cell of the developed sensor revealed a characteristic dipole-like pattern, with pronounced field localization at the ends of the graphene microstrips, indicating strong localization. This field enhancement near the sensing interface facilitates efficient interaction with the analyte layer, thereby improving the sensitivity of the sensor.

The results of the current study demonstrate that the designed metasurface-based sensing structure with graphene microstrips can be applied as a sensor for non-destructive determination of albumin and other biomolecules in liquid samples.

## ACKNOWLEDGMENT

This work was supported by grant № 0122U001687 of the NAS of Ukraine; K. S. Kuznetsova acknowledges the support provided by "The Pauli Ukraine Project", funded by the Wolfgang Pauli Institute Thematic Program "Mathematics-Magnetism-Materials" (2023/2024); Z. E. Eremenko acknowledges the funding from the European Union under the Marie Skłodowska-Curie grant agreement no. MSCA4Ukraine project number 1.4 - UKR - 1232611 - MSCA4Ukraine (IFW Dresden). 1.4 - UKR - 1232611 project has received funding through the MSCA4Ukraine project, which is funded by the European Union. Views and opinions expressed are however those of the author(s) only and do not necessarily reflect those of the European Union, the European Research Executive Agency or the MSCA4Ukraine Consortium. Neither the European Union nor the European Research Executive Agency, nor the MSCA4Ukraine Consortium as a whole nor any individual member institutions of the MSCA4Ukraine Consortium can be held responsible for them.

# Терагерцовий сенсор на основі метаповерхні з графеновими мікросмужками для біодетектування: моделювання та застосування

К. С. Кузнецова[1], В. А. Пашинська[1,2], З. Є. Єременко[1,3]

[1] Інститут радіофізики та електроніки ім. О.Я. Усикова НАН України, вул. Академіка Проскури, 12, Харків 61085, Україна, *kuznetsova.ire@gmail.com*

[2] Фізико-технічний інститут низьких температур ім. Б.І. Вєркіна НАН України, пр. Науки, 47, Харків, 61103, Україна, *vlada.pashynska@gmail.com*

[3] Інститут дослідження твердого тіла та матеріалів ім.. Лейбніца, 01069 Дрезден, Німеччина, *zoya.eremenko@gmail.com*

Дане дослідження методами чисельного моделювання присвячене розробці та оптимізації сенсора на основі метаповерхні з графеновими компонентами для потенційних застосувань в галузі біодетекції. Елементарна комірка запропонованої метаповерхні складається з тонкого гнучкого діелектричного шару підкладки з центрально розташованою графеновою мікросмужкою. За результатами чисельного моделювання спектральних властивостей метаповерхні за допомогою програмного забезпечення COMSOL Multiphysics у терагерцовому діапазоні від 5 до 35 ТГц виявлено максимуми спектру поглинання (резонансні моди) при $f_1$ = 8,7 ТГц та $f_2$ = 26,5. Структурні параметри розробленої метаповерхні з графеновими мікросмужками були налаштовані для досягнення оптимальних резонансних властивостей. Наступні етапи дослідження демонструють, що розміщення шару рідкого зразка, що тестується (води або розчину бичачого сироваткового альбуміну (БСА), на метаповерхні викликає низькочастотний зсув плазмонної резонансної моди $f_1$, обраної для біосенсорних вимірювань. Цей зсув частоти, разом зі зміною амплітуди піку поглинання, демонструють чутливість до показника заломлення рідкого зразка. Резонансна поведінка розробленої структури метаповерхні визначається збудженням локалізованого плазмонного резонансу в графенових елементах та ефектом електромагнітного зв'язку ближнього поля між короткими краями графенових мікросмужок. Оцінка впливу матеріалу діелектричної підкладки на чутливість сенсора на основі метаповерхні до змін концентрації БСА показує, що підкладка Kapton забезпечує вищу ефективність порівняно з підкладкою з SiC. Отримані результати демонструють потенціал розробленого сенсора на основі метаповерхні з графеновими мікросмужками для застосування як сенсорної структури для визначення білків та інших біомолекул у рідких зразках.

Ключові слова: метаповерхня, графенові мікросмужки, біосенсор, спектр поглинання, терагерцовий діапазон, бичачий сироватковий альбумін, чисельне моделювання.